\documentclass{article}

\usepackage[T1]{fontenc}

\usepackage[margin=1.1in]{geometry}

\usepackage{newtxtext} % change text font to Times New Roman

\usepackage{xargs}    % Use more than one optional parameter in a new commands

\usepackage{amsmath,amssymb,amsthm}
\usepackage{mathrsfs}
\usepackage{derivative}
\usepackage{mathtools,bm}

\usepackage{titlesec}
\usepackage{authblk}

\usepackage{hyperref}
\usepackage{cleveref}

\usepackage[numbers,sort&compress]{natbib}
\usepackage{graphicx,epstopdf, epsfig}
\usepackage{subcaption}
\usepackage{rotating}
\graphicspath{{./figs/}}       % add path to figures directory

\titleformat{\paragraph}
{\normalfont\normalsize\bfseries}{\theparagraph}{1em}{}
\titlespacing*{\paragraph}
{0pt}{3.25ex plus 1ex minus .2ex}{1.5ex plus .2ex}

\begin{document}

\title{Volume filtered FEM-DEM framework for simulating particle-laden flows in complex geometries}
\author[1]{Abhilash Reddy Malipeddi\thanks{Corresponding author. \mbox{e-mail:~absh@umich.edu}}}
\author[2,3]{C. Alberto Figueroa}
\author[1,4]{Jesse Capecelatro}
\affil[1]{\footnotesize Department of Mechanical Engineering, University of Michigan, Ann Arbor, MI 48109, USA.}
\affil[2]{\footnotesize Department of Biomedical Engineering, University of Michigan, Ann Arbor, MI 48109, USA.}
\affil[3]{\footnotesize Department of Surgery, University of Michigan, Ann Arbor, MI 48109, USA.}
\affil[4]{\footnotesize Department of Aerospace Engineering, University of Michigan, Ann Arbor, MI 48109, USA.}
\date{}
\maketitle
\vspace{-2em}
\begin{abstract}
We present a computational framework for modeling large-scale particle-laden flows in complex domains with the goal of enabling simulations in medical-image derived patient specific geometries. The framework is based on a volume-filtered Eulerian-Lagrangian strategy that uses a finite element method (FEM) to solve for the fluid phase coupled with a discrete element method (DEM) for the particle phase, with varying levels of coupling between the phases. The fluid phase is solved on a three-dimensional unstructured grid using a stabilized FEM. The particle phase is modeled as rigid spheres and their motion is calculated according to Newton's second law for translation and rotation. We propose an efficient and conservative particle-fluid coupling scheme compatible with the FEM basis that enables convergence under grid refinement of the two-way coupling terms. Efficient algorithms for neighbor detection for particle-particle collision and particle-wall collisions are adopted. The method is applied to a few different test cases and the results are analyzed qualitatively. The results demonstrate the capabilities of the implementation and the potential of the method for simulating large-scale particle-laden flows in complex geometries.
\end{abstract}

\section{Introduction}
Fluid mechanics plays a crucial role in many physiological processes on health and disease. Given recent advances in medical imaging, computational power, and mathematical algorithms, real-time patient-specific computational fluid dynamics is now becoming possible. We present a computational framework capable of capturing fluid-particle interactions in complex medical-image derived geometries that are much larger than the size of the particles and with large number of particles. A four-way coupled volume-filtered Euler-Lagrange solver with a stabilized finite element based incompressible flow solver is developed.

Most of the development around particle-laden flows has been in the context of industrial applications such as gas-solid flows in fluidized beds using structured grids. 
In contrast, the application areas of interest here are biological flows which often involve complex organic geometry.
Finite element method based incompressible flow solvers have been gaining popularity since the development of stabilization schemes. They are attractive for their nice mathematical properties and their ability to naturally deal with complex geometries. FEM based solvers for Euler-Lagrange type particle-laden flows are an active area of research. \citet{casagrandeHybridFEMDEMApproach2017} have developed a stabilized FEM based incompressible solver for particle-laden and porous flows that was restricted to structured grids. Recently \citet{elgeitaniHighOrderCFDDEMDevelopment2023b} have developed a FEM based particle-laden flow solver for gas-particle flows in engineering capable of handling unstructured hexahedral meshes for industrial applications.
\citet{Dufresne2020} have implemented and demonstrated a massively parallel particle-laden flow solver in the \texttt{YALES2} code. This uses a finite volume discretization for the fluid phase.
While some ``simple" complex boundaries can be represented on structured grids through immersed boundary methods for example, this is in general not very efficient and the fidelity of the representation is limited by the grid size.
Presence of large empty regions in the domain can lead to excessive memory usage and computational cost due to unused grid points. Unstructured grids are better because they capture the complex geometry more accurately in comparison. But, unstructured grids are more challenging to work with. The development of efficient algorithms for particle tracking on unstructured grids is an active area of research. 

In this work we present the development of a versatile and massively parallel framework for studying biological particles in subject-specific geometries by combining (i) a recently-developed statistical hydrodynamic model for fluid- particle flows; (ii) scalable Eulerian–Lagrangian algorithms; and (iii) state-of-the-art parallel techniques for simulating biological flows in image-based geometrical models. 
There are few salient aspects that set the current work apart. The inter-phase coupling procedure developed here can be applied to a wide range of particle and mesh sizes at once. It is shown to be highly efficient, consistent and convergent. The particle-wall collision algorithm is more efficient and avoid preprocessing steps.
First we briefly describe the mathematical model and particularly the volume-filtering procedure as applied to the particle-fluid. Then we describe algorithmic developments that enable scalable tracking of large number of particles within unstructured grids. Next we describe the collision processing for inter-particle and particle-wall collisions that can efficiently deal with complex non-convex boundary collisions. Next, we describe a novel two stage particle-fluid interphase-coupling transfer function that is efficient, conservative and convergent on unstructured grids.

\section{Volume-filtered Euler-Lagrange equations}
The volume-filtered Euler-Lagrange approach to modeling particle-laden flows is described in this section. The general idea is that the particle phase is treated as being composed of Lagrangian particles with properties such has diameter, density, velocity and potentially other scalar quantities like temperature or species concentration. The effect of the fluid on the particles is modeled using a so-called {\em drag law}, whereas the particle-particle collisions are captured explicitly. This approach offers a good compromise between expensive particle-resolved simulation methods and two-fluid/Eulerian-Eulerian methods that rely heavily on sub-grid models to account for both particle-particle and particle-fluid interactions. The equations for the fluid phase and particle phase and the coupling is described next.

\begin{figure}
    \centering
    \includegraphics[width=0.45\textwidth]{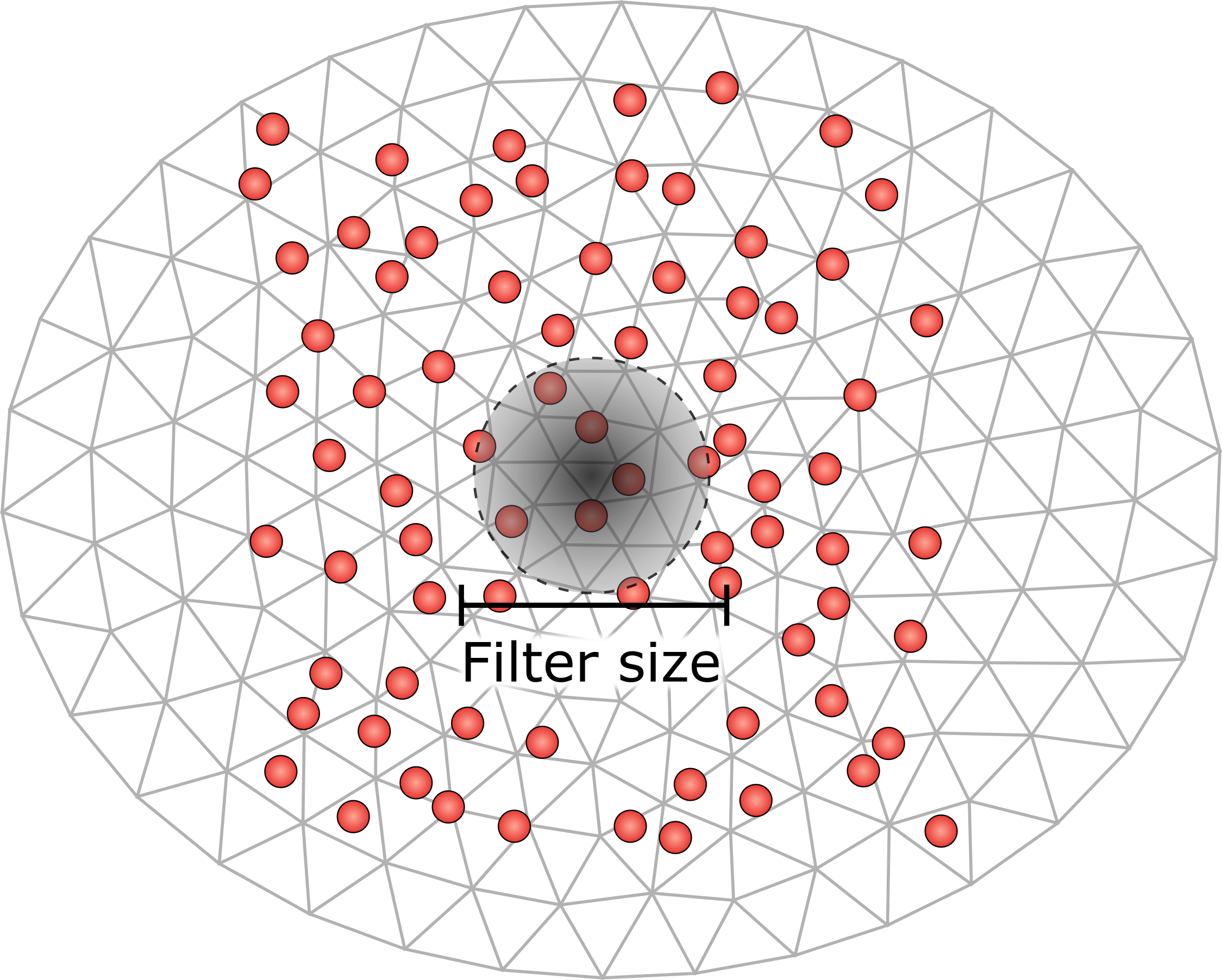}
    \caption{Schematic of Lagrangian particles inside a discretized fluid domain along with a representation of the filtering kernel.\label{fig:vfel}}
\end{figure}

\subsection{Fluid-phase equations}
The equation for the fluid phase are obtained by applying a spatial filtering operator on the incompressible Navier-Stokes equations, taking into account the volume occupied by the particles and the momentum exchange between the particles and fluid phase. The detailed derivation of the equations can be found in \cite{andersonFluidMechanicalDescription1967a,Capecelatro2013}. \Cref{fig:vfel} illustrates the filtering procedure on a system of particles suspended in a fluid. The resulting continuity equation after filtering reads
\begin{equation}
    \frac{\partial \phi_f}{\partial t} + \frac{\partial}{\partial x_i}\left(\phi_f u_i\right) = 0
\end{equation}
and the momentum equation is

\begin{equation}
    \rho\phi_f \frac{\partial u_i }{\partial t} + \rho\phi_f u_j\frac{\partial u_i}{\partial x_j}=  -\phi_f\frac{\partial p}{\partial x_i} +\phi_f\frac{\partial \tau_{ij}}{\partial x_j} + \phi_f\rho g_i + f^p_i
\end{equation}
where $\tau_{ij} = \mu\left(u_{i,j}+u_{j,i}\right) $, $\mu$ is the fluid viscosity, $\rho_f$ is the fluid density, $g_i$ is the body force. The term $f^p_i$ is the momentum feedback force from the particles acting on the fluid. $\phi_f$ is the volume fraction of the fluid phase. 
The model used to calculate the force on the particle due to the fluid is described next.
%This can be written as
% \begin{equation}
    % f^p_i = -\sum_k^N \frac{f^{h,k}_i}{V_p}
% \end{equation}
% where $V_p$ is the volume of the particle and $f^{h,k}_i$ is the hydrodynamic force on the \emph{k}\textsuperscript{th} particle due to the fluid.
%The fluid equations are solved on a three dimensional unstructured grid using a stabilized finite element method.

\subsection{Particle equations}
The particles are modeled as rigid spheres.
Their motion is calculated according to Newtons second law for translation and rotation.
For spherical particles, due to the symmetry, tracking the absolute orientation is not strictly necessary.
The rotational velocity is required to calculate the hydrodynamic torque based on the difference in the particle and fluid rotational velocity. The translational motion for particle $p$ is given by
\begin{equation}
    m_p\odv{v_i}{t} = f_i^h + \sum_a f^{col}_{i,p \leftarrow a} + (\rho_p-\rho_f)V_p g_i
\end{equation}
where $m_p$ is the mass of the particle, $V_p$ is the volume of the particle, $f_i^h$ is the hydrodynamic force exerted on the particle by the fluid (elaborated further down).
$f^{col}_{i,p \leftarrow a}$ is the force on the particle due to collisions with other particles (and walls).
Within the summation, $p \leftarrow a$ denotes the collision of particle $p$ with particle $a$.
Gravity is included as a body force term denoted with the acceleration due to gravity $g_i$.
Since we do not include the resolved fluid stress force in the particle momentum equation, we include a buoyancy force that takes into account the difference in the density of the particle ($\rho_p$) and the fluid ($\rho_f$).
In the biofluids application area $\rho_p/\rho_f \sim \mathcal{O}(1)$.
If the resolved fluid stress force is included in the particle momentum equation, then we would need to add the total gravitational force on the particle, $\rho_p V_p g_i$ to the right hand side of the particle momentum equation.

The angular momentum equation conservation for the \emph{p}\textsuperscript{th} particle is given by
\begin{equation}
    I\odv{\omega_i}{t} = \tau^d_i + \sum_a\frac{d_p}{2} \epsilon_{ijk}n_j f^{tcol}_{k,p \leftarrow a}
\end{equation}
where $I$ is the moment of inertia, $\omega_i$ is the angular velocity of the particle. The first term on the right hand side is the hydrodynamic torque due to drag and the second term the summation of the forces due to the tangential component of the frictional force due to collisions.
The torque acting on a small isolated sphere rotating in a quiescent viscous fluid is given by
\begin{equation}
    \tau^d_i = \pi\mu d^3_p\left(\omega^f_i-\omega^p_i\right)
\end{equation}
where $\omega^f_i$ is the angular velocity of the fluid interpolated to the particle center and $\omega^p_i$ is the angular velocity of the particle.

The hydrodynamic force can be written as a superposition of forces arising from distinct mechanisms.
\begin{equation}
    f_i^h = f_i^d + f_i^a + f_i^b + f_i^l+ f_i^m
\end{equation}
where $f_i^d$ is the quasi-steady drag force, $f_i^a$ is the added mass force due to the acceleration of the fluid around the particle, $f_i^b$ is the Basset history force to account for viscous effects due to unsteady motion of the particle, $f_i^l$ is the Saffman lift force due to the pressure distribution that develops on the surface of a particle in velocity field with a non-zero gradient, and finally the Magnus lift force $f^m$ due to the rotation of the particle.
In general, this linear superposition of individually identified hydrodynamic forces on the particle is not well-founded, but is invariably used in literature.
It can be shown to hold in the low and high Reynolds number limits.
In the absence of better alternatives, we do the same and superpose the individually identified hydrodynamic forces on the particle.

For the application areas of interest, Basset force and Magnus force are not expected to be significant and are not included in the simulations.
In some Euler-Lagrange formulations in literature, there is an additional fluid stress force (resolved stress term) added to the RHS. The particular Euler-Lagrange formulation used here accounts for the resolved fluid force in the fluid momentum equation and hence this term is not included in the particle momentum equation.
See \citet{Zwick2020} for a detailed discussion on the different formulations and their equivalence.

The added mass force is given by
\begin{equation}
    f_i^a = \frac{1}{2}\rho V_p \left( \mdv{u_i}{t}  - \odv{v_i}{t}\right)
\end{equation}
where $ \mdv{~}{t}$ is the material derivative and $v_i$ is the velocity of the particle. There are alternative forms in literature that account for the finite size of the particle by adding a term called Faxen's correction, that has not been included here. $\odv{~}{t}$ is the time derivative in the particle frame of reference. The Saffman lift force is calculated according to the expression given by \citet{mclaughlinLiftSmallSphere1993}. The equations are integrated using a second order accurate scheme. The fluid solver is fully implicit and as a result can take large time steps. If the fluid time step is larger than the particle's stable time step, the particle equations are sub-cycled in time to avoid stability issues.

\subsubsection*{Quasi-steady drag force}
The quasi-steady drag force is calculated using the relation provided by \citet{Tavanashad2021} for freely evolving suspensions of particles.
The correlation was developed from particle-resolved direct numerical simulations conducted for a range of Reynolds numbers, volume fractions, and density ratios.
Most of the prior work on developing the drag relations assumed particles to be much denser than the fluid phase, which is typically the case in gas-solid flows relevant to many industrial applications such fluidized bed reactors for example.
Consequently, most of the the particle-resolved simulations consider fixed particle packing and the particles are not allowed to move.
For the application being considered here, i.e., biological flows, the difference in the density of the particles and the fluid is not very large and the assumption of fixed particle packing might reduce the accuracy of the drag relations. 
For $\rho_p/\rho_f <10$ the formula for the drag force is
\begin{equation}
    f^d_i\left(\phi, \operatorname{Re}_m\right)=f^{St}_i \left( 1 + 0.15\operatorname{Re}_m^{0.687}\right) \left(78.96\phi_p^3 - 18.63\phi_p^2 + 9.845\phi_p + 1\right)
\end{equation}
where $f^{St}_i = 3\pi\mu d_p\left(u_i^f-v_i\right)$ is the Stokes drag force and $\phi_p$ is the volume fraction of the particles. The modified Reynolds number is defined as
\begin{equation}
    \operatorname{Re}_m = \frac{\rho_f d_p\phi_f \left|u_i^f-v_i\right|}{\mu}
\end{equation}

\section{Numerics}
\subsection{Fluid discretization}
The fluid phase equations equations are solved using the finite element method.
The filtered equations are written in their weak form and discretized using the finite element method. The mass and momentum conservations can be rearranged into a convenient form and written respectively as
\begin{align*}
    u_{i,i} + \frac{1}{\phi_f}\left(\phi_{f,t} + u_j\phi_{f,j}\right)                             & = 0,   \\
    \mathscr{L}_i = \rho u_{i,t} + \rho u_{j}u_{i,j} + p_{,i} -\tau_{ij,j} - \frac{f^p_i}{\phi_f} & = 0_i. \\
\end{align*}

The weak form of the above equations is obtained by multiplying by test functions and integrating over the domain
\begin{align*}
    \int_{\Omega} \left\{  qu_{i,i} + \frac{q}{\phi_f}\left(\phi_{f,t} + u_j\phi_{f,j}\right) + w_i~\left( \rho u_{i,t} + \rho u_{j}u_{i,j} + p_{,i} -\tau_{ij,j} - \frac{1}{\phi_f}f^p_i \right) \right\} ~\mathrm{d}\Omega = 0,
\end{align*}
where $q$ is the test function for the divergence constraint and $w_i$ is the test function for the momentum equation.
This is the standard (meaning unstabilized) Galerkin weak formulation. We will refer to these terms as Galerkin terms (to separate them from the stabilization terms that we will add below). This is formulation is not stable in general. There are two issues to overcome. One is the spurious upstream disturbance at high $Re$ and the other is spurious modes in the pressure for certain combination of velocity-pressure elements related to the LBB or  inf-sup condition. Same type of elements for velocity and pressure in particular do not work. There is some attraction to use same element types for both velocity and pressure.
To remedy the high $Re$ problem and to remove restrictions on the choice solution function spaces we add stabilization terms to the formulation. Let us denote the last LHS by $B_G(w_i,q;u_i,p)$. With the stabilization terms added the formulation is
\begin{align*}
    B(w_i,q;~u_i,p) & = 0                                                                                                                                                                                                                                                                        \\
    B(w_i,q;~u_i,p) & = B_G(w_i,q;~u_i,p)                                                                                                                                                                                                                                                        \\
                    & + \sum_{e=1}^{n_{el}} \int_{\bar{\Omega}_e}\left\{ \tau_M(u_jw_{i,j}+\frac{q_i}{\rho})\mathscr{L}_i + \tau_Cw_{i,i}u_{j,j}\right\}~\mathrm{d}\bar{\Omega}_e                                                                                                                \\
                    & + \sum_{e=1}^{n_{el}} \int_{\bar{\Omega}_e}\left\{ w_i \rho\overset{\Delta}{u_{j}} u_{i,j} + \bar{\tau} \left\{\frac{\rho\overset{\Delta}{u_{j}}}{\tau_M}\right\}w_{i,j}\left\{\frac{\rho\overset{\Delta}{u_{k}}}{\tau_M}\right\}u_{i,k} \right\}~\mathrm{d}\bar{\Omega}_e, \\
\end{align*}
The details of the stabilization scheme and the complete details of the numerical approach can be found in \cite{whitingStabilizedFiniteElement2001}. In brief, consistent terms are added to the weak form to stabilize the system.
In the continuous limit, the stabilization terms vanish and we are left with a consistent set of equations.
The non-linear fluid equations are solved implicitly in a monolithic fashion using a Newton-Raphson method. The time integration scheme used is second order generalized-$\alpha$ method suitably modified for the system of differential algebraic equations that arise from the discretization of the fluid equations. We use linear tetrahedral elements for all variables.

The implementation is done within the \texttt{CRIMSON} framework \cite{ArthursPLOS_2021}. \texttt{CRIMSON} specializes in simulation of blood flow in the vasculature. It has advanced capabilities to read-in medical images and allows the user to create a corresponding geometry and mesh and apply complex physiologically accurate dynamic boundary conditions. The underlying flowsolver is a general purpose finite element solver, PHASTA \cite{whitingStabilizedFiniteElement2001} that solves the incompressible Navier-Stokes equations on unstructured grids. The solver is parallelized using MPI and has been demonstrated to scale to 100,000 cores\cite{Zhou2010}.

\subsection{Particle discretization}
The particle related computations are described in this section. First we describe how the particles are tracked through the domain as they are carried by the flow. Then we describe how the collisions are detected and processed.

\subsubsection{Particle tracking}

Tracking particles in an unstructured grid is a challenging task. The main difficulty is that the elements are not arranged in a regular fashion. So, we cannot use a simple indexing scheme to identify the element that contains the particle. During the course of the simulation, as teh particles are advected by the flow, the particles will move from one element to another. After each particle has been advected, we identify the new element host for each particle by executing a delaunay search with the last known element as the starting point. For the applications of interest, the time-step is small enough that the particles do not move very far and the delaunay search works well.

When the particles are initialized, as a preprocessing step we identify and record the starting element for each particle to enable the quick delaunay search.

The particle tracking and collision detection is implemented using a two stage approach. The first stage is to identify the element that contains the particle. The second stage is to localize the particle within the element. The first stage is implemented using a cell list method. The second stage is implemented using a delaunay search method.
In prior work \citet{Rydquist2020} have used a cell list method for particle tracking. The idea is to overlay a Cartesian grid on top of the unstructured mesh and assign each cartesian cell with all the unstructured mesh elements that intersect with this cell.

% This is done in a preprocessing step. Then, for a given query point, the Cartesian cell that contains the point is identified and the element that was assigned to this cell is used as the starting element for the delaunay search procedure. This method is simple and efficient. The only drawback is that it requires a preprocessing step. This is not a problem for static meshes. But for moving meshes, the preprocessing step needs to be repeated every time the mesh is updated. This is not a problem for the application being considered here because the mesh is static. The cell list method is described in detail in the next.

First, we initialize a integer array for the background grid. We will store just one value per cell. This can be a hashmap if the domain has a lot of empty space and memory usage is a concern. Lets say that the side of the cell is approximately twice as large as a characteristic length of the largest element in the unstructured mesh.

Then, we loop over the elements and for each element find indices of the cell that contains the first vertex of this element. If this cell has not already been assigned, assign this element id to this cell. At this point one might think that that all relevant cells have been assigned a value. If we draw the cells that have been assigned overlaid on the mesh, we would like to see that the cartesian cells cover the entire region. But it is very likely that there are cells which did not have the vertex that was tested( or any vertex) inside them. So, there will be "holes" if the cartesian cells are too small. Additionally, near the boundaries (particularly curved or inclined) it is possible that there are parts of the element that do not have a cartesian cell covering it.
If the cartesian cells are not too small ( i.e. large enough that each cell contains at least one whole element), then the holes do not appear. To deal with the potential lack of coverage with boundary elements, the following is done.
We loop over the cartesian cells and for each cell check if the immediate neighbors are assigned. if not we make a note of the indices and the value that will be assigned to this adjacent cell will be the value assigned this cartesian cell. This needs to be done in two stages, identification and assignment, because we wouldn't know if a cell was assigned the first time or during this step. The effect will be that we are padding the cartesian cells with one layer of cells. This ensures full coverage of boundary elements.

\subsubsection{Collision processing}
This section described the algorithms adopted for scalable calculation of particle-particle and particle-wall collisions are described. It is well known that a naive implementation of nearest neighbor search for particle-particle collisions scales as $\mathcal{O}(N_p^2)$. Similarly a naive implementation of particle-wall collisions scales as $\mathcal{O}(N_p N_w)$ where $N_w$ is the number of wall elements. This is a non-starter even for a moderate number of particles. To enable simulation of \(\mathcal{O}(10^9)\) particles we need to avoid the quadratic and multiplicative scaling of the collision detection algorithms. Such algorithms for both particle-particle and particle-wall collisions are described below.

\paragraph{Particle-particle collisions}
\begin{figure}
    \centering
    \includegraphics[width=0.5\textwidth]{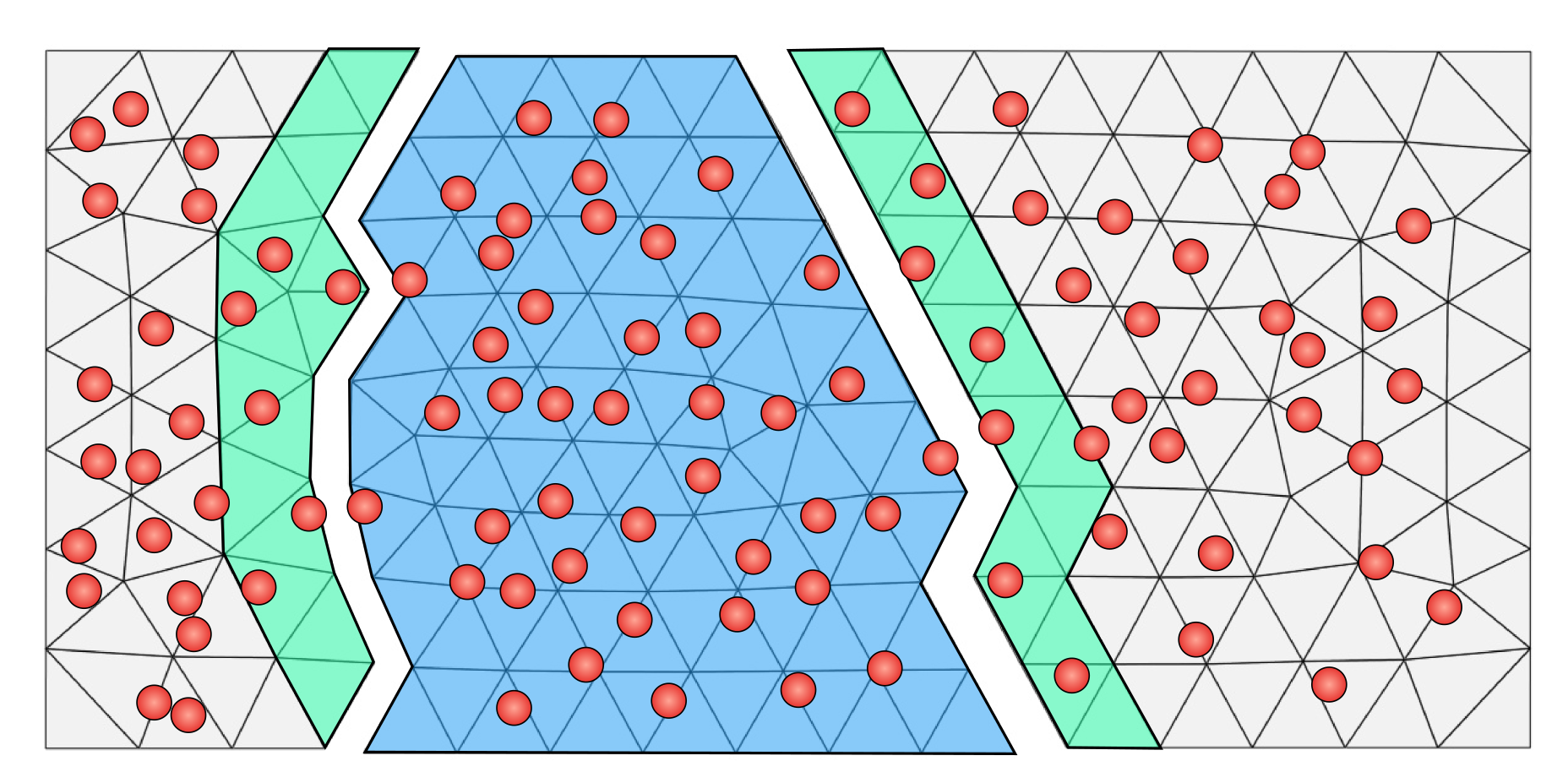}
    \caption{Using a distance field to the partition boundary, we identify particles on neighboring processes that might collide with particles on a given process and communicate these ghost particles, the particles that are in the green shaded region, to the process that owns the blue shaded partition.\label{fig:coll_par}}
\end{figure}
The flow solver uses domain decomposition to solve the equations in parallel. Working within this framework, we store the particles on the process that contains the element that contains the center of the particle.
A simple example of a meshed domain that has been decomposed into three partitions is shown in \cref{fig:coll_par}.
Let us consider the partition highlighted in blue. To capture all the collisions of the particles in this partition, the process that owns this partition needs to know about particles on neighboring processes that are nearby the shared partition boundary within some cut-off distance.
This distance will depend on the size of the particles. As we use a soft sphere model for collisions, the cut-off distance should be at least the diameter of the largest particle.
If there are some guarantees about the relative size of the particles with respect to the mesh size, then, the  elements in the cut off region could be counted based on their {\em graph distance}, i.e. we could determine this halo zone in terms of layers of cells.
But in general such assumptions are not possible and a general approach is called for. Here, we identify the the elements in the cut off region by calculating the distance field from the from the partition boundary to the interior of the partition.
Then, the particles in the cut off region are identified by interpolating the distance field to the particle centers. These are then communicated to the corresponding neighboring process. This is illustrated in \cref{fig:coll_par} where the elements highlighted in green are the elements that contain particles that can potentially collide with particles in the blue partition.

\begin{figure}
    \centering
    \includegraphics[width=0.35\textwidth]{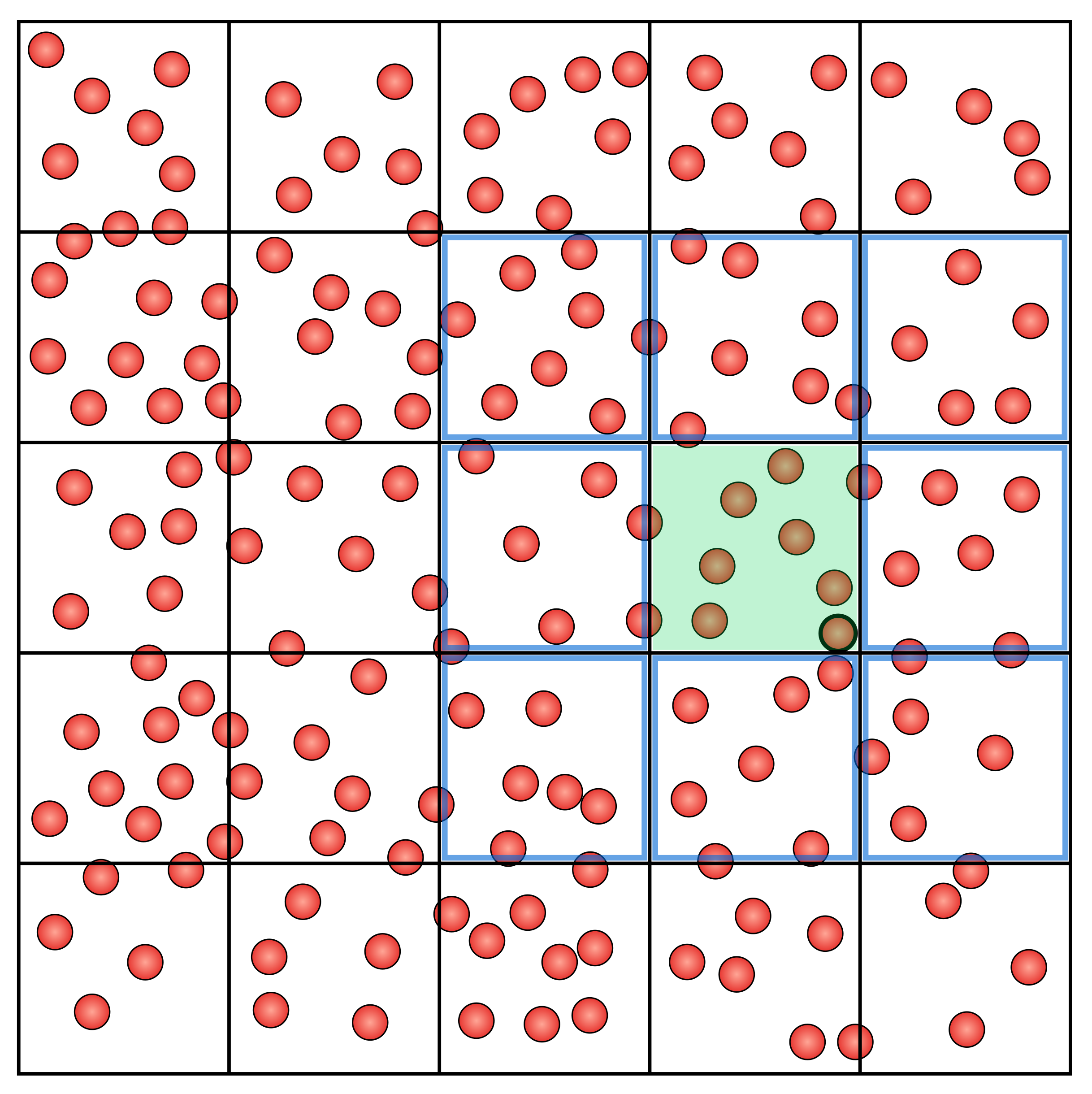}
    \caption{After ghost particles are communicated, a cell list is used to identify nearby particles for collision detection. To identify potential collisions for particles in the green cell, we only need to check the particles in the green cell and the neighboring blue-outlined cells.\label{fig:cell_list}}
\end{figure}

\begin{figure}
    \centering
    \includegraphics[width=0.5\textwidth]{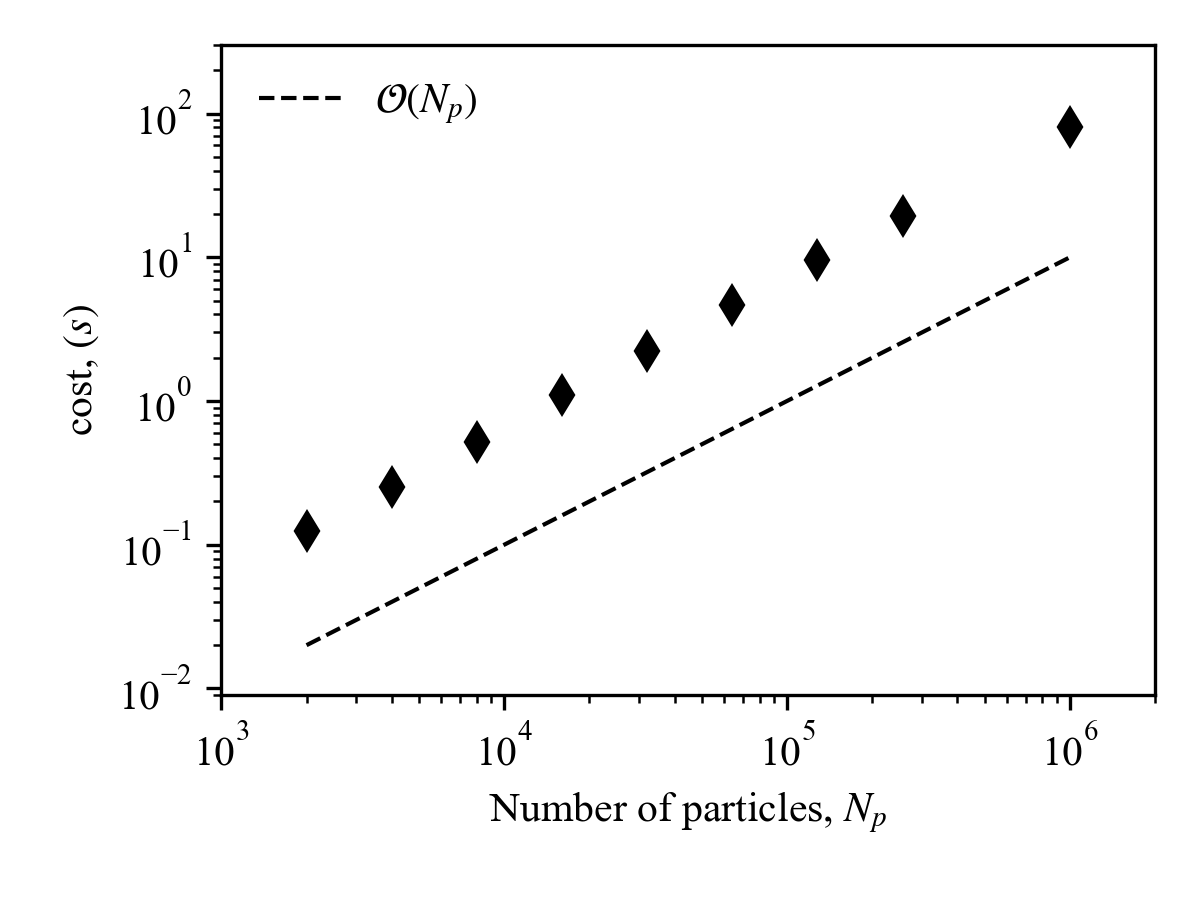}
    \caption{Scaling of the collision detection algorithm with number of particles\label{fig:coll_scaling}}
\end{figure}

Once all the particles are available on a process, the remaining work is processor local and does not depend on parallelization details. Using a cell list we divide the bounding box containing the particles into Cartesian cubes. The size of the cube is based on the largest particle diameter. We loop over the particles and assign them to the cell that contains the particle center. Once this is done for all particles, to identify collisions for a given particle, we only need to check the particles in the cell that contains the particle and the neighboring cells ($1 + 26=27$ neighbors in 3D). This is illustrated in \cref{fig:cell_list} in two dimensions. To compute all possible collisions for a particle in the green cell, we only need to check the particles in the green cell and the neighboring blue-outlined cells. This is a significant reduction in the number of particles that need to be checked for collisions. The scaling of the algorithm with the number of particles is shown in \cref{fig:coll_scaling}.
The choice of the size of the cell list box is important. For optimal results, \citet{Rydquist2020} has shown that the number of cells should approximately be equal to the number of particles.

Since we are dealing with complex tortuous geometries, the distribution of the particles within the bounding box will not be uniform. This can lead to excessive memory usage if we allocate a 3d array of lists to store the particles in each cell. To avoid this issue, we use a locality sensitive hashing scheme to store the particles in a unsorted key-value store. The key is a hash of the cell indices and the value is a list of particles. This is implemented using the \texttt{std::unordered\_map} data structure in C++, also known as a {\em hash map}. With this method, if a cell is empty, we do not need to allocate any memory for it. This is a significant advantage over the 3d array of lists approach. The particular hash function used is the {\em 3D Morton index} or {\em Z-order curve}. This is a space filling curve that reversibly maps a 3D point to a 1D index. The mapping is such that points that are close in 3D space are mapped to indices that are close in 1D space. This is a desirable property for the hash function because it means that the particles that are close in 3D space will be mapped to keys that are close in 1D space. This will lead to a more uniform distribution of the particles in the key-value store. In addition, since this is a reversible hash, by knowing the key, we can recover the 3D cell index for any particle. The scaling of this collision detection algorithm with the number of particles is shown in \cref{fig:coll_scaling} going up to a million particles. We get the optimal linear scaling with the number of particles.

\begin{figure}
    \centering
    \includegraphics[width=0.3\textwidth, angle=90,origin=c]{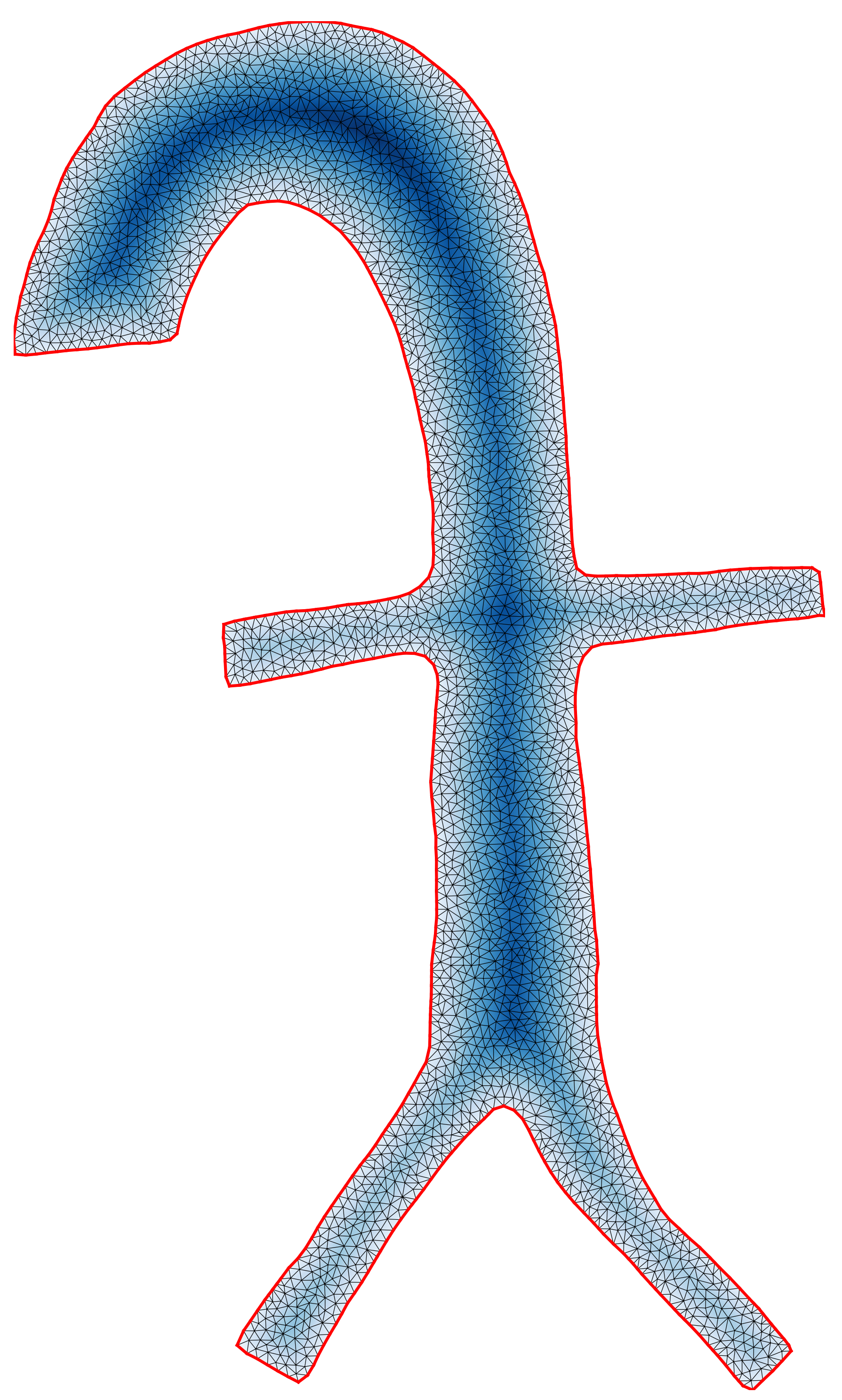}
    \vspace{-4em}
    \caption{Contour plot of the boundary distance field for a 2D aorta geometry. The farther away from the boundary the darker the color.\label{fig:distfield}}
\end{figure}

\paragraph{Particle-wall collisions}
The collision detection can be accelerated if we only check the boundary collisions of only those particles that are near the boundaries and ignore the particles in the interior.
To enable quick identification of near wall particles, we initialize a distance field due to the boundary of the domain in the interior of the domain. At any point in the interior, the distance field is the shortest distance to the nearest boundary. This is illustrated in \cref{fig:distfield} for a 2D aorta geometry.
Near wall particles are identified by interpolating the distance field to the particle center and checking if the distance is less than the about 1.2 times radius (including a factor of safety).
Once the near boundary particles have been flagged, we use a another cell-list like structure and assign the boundary triangles to the cells. This reduces the number of boundary features that need to be checked to only those that are nearby the particle and skips triangles, edges and vertices that are far away.
Then, for each particle we easily identify the cell and test for collision with the boundary faces, edges and vertices in the neighboring cells and calculate the collision force.
We first calculate a distance field in the interior of the domain. To quickly eliminate particles in the interior, we use the FEM interpolating functions to calculate the distance at the particle centers.

\begin{figure}
    \centering
    \includegraphics[width=0.65\textwidth]{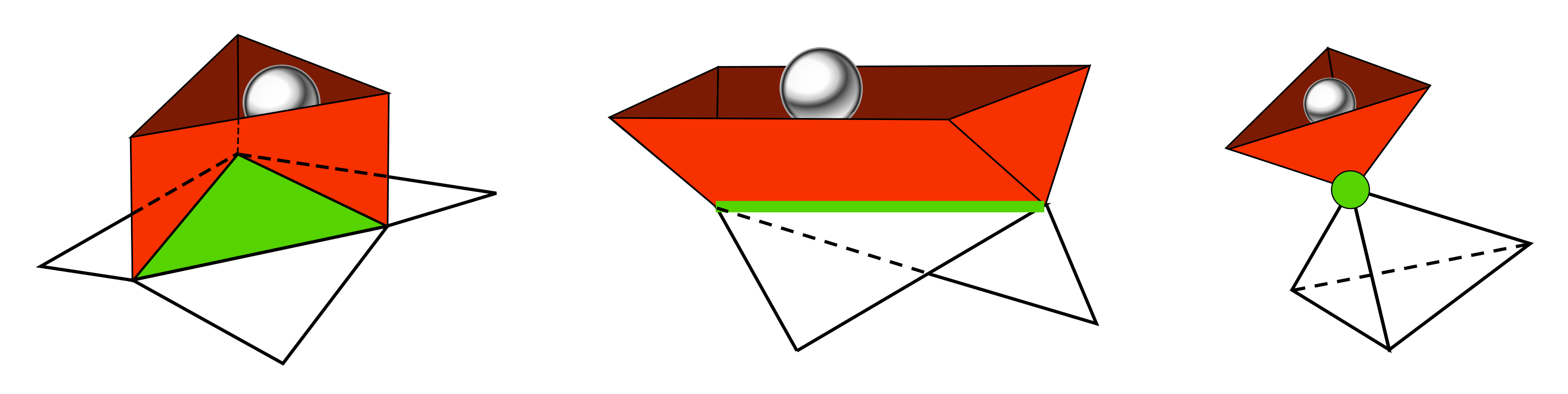}
    \caption{Voronoi region near the domain boundary for three different types of geometric features that make up the boundary, i.e., faces, edges and vertices.\label{fig:voronoi_regions}}
\end{figure}

In general, the particles near the domain boundary could collide with a boundary feature such as a face, an edge, a vertex or any permutation of multiple of them. Robust and unambiguous processing of collisions is made possible by using the concept of {\em Voronoi regions}\cite{Dufresne2020}. The near boundary region is divided into regions based on which boundary feature is nearest to the point in the region.
This is illustrated in \cref{fig:voronoi_regions}.
Particles near the boundary are assigned to the region corresponding to the feature that is nearest to the particle center. Collision force is calculated only if the particle is in the Voronoi region of the boundary feature it is intersecting with.
The algorithm adopted here first checks for face collisions. These are always valid and are registered immediately. For any faces that are registered, we void the edges and vertices. That is, we ignore any intersection of this particle with any edge or vertex that were part of previously registered faces. Then, we do a pass over the edges and register edge collisions and void the vertices for the edges that are registered. Finally, we do a pass over the vertices and register vertex collisions.

The approach presented here deals with correctly with all possible collision scenarios regardless of the local curvature characteristics of the boundary surface. It is a mathematically exact and exhaustive procedure without any user-specified tolerance that might require tuning. In literature, the idea of Voronoi region has been used but it required an expensive preprocessing step to actually build and store the Voronoi information. The algorithm described here implicitly implements the Voronoi region idea and does not require a preprocessing step and has improved efficiency while retaining the robustness and unambiguous nature of the original Voronoi region approach.

To calculate the actual collision force, once the collision normal and penetration depth are calculated, we treat the wall as a particle with infinite mass and calculate the collision force as if it were a particle-particle collision using the same soft sphere model.

\section{Particle-fluid coupling}
The coupling between particles and fluid is a crucial aspect of Euler-Lagrange simulations. Depending on the specific characteristics of the particle-laden flow being simulated, different levels of physical interactions between the particle and fluid phases can be incorporated in the simulation.
In one-way coupling, the fluid exerts a drag force on the particles, but the particles do not affect the fluid. This is typically observed in dilute flows where the particle volume fraction is small and the particles have minimal impact on the fluid flow.
Two-way coupling involves interactions in both directions, where the particles affect the fluid and the fluid affects the particles. This is commonly seen in dense flows where the particles significantly influence the fluid flow and vice versa.
Four-way coupling takes into account not only the interactions between particles and the fluid, but also considers inter-particle collisions. This is suitable for dense flows with strong particle-particle and particle-wall collisions.
Within two-way and four-way coupling, it is possible to ignore the excluded volume effects of the particles if the particle volume fraction is small, but still consider the momentum exchange between the particles and the fluid.

In one-way coupled simulations, we need to compute the fluid velocity, vorticity, fluid stresses etc. at the particle center to be able to calculate all the hydrodynamic forces on the particle and integrate the particle equations.
For higher levels of coupling in addition to the Eulerian to Lagrangian transfer, we also need to have Eulerian representations of particle centered quantities, mainly the volume fraction field and feedback force.

Ideally, transfers in both directions would be done consistently using the same filtering kernel function by identifying the support of the kernel function and integrating over the support. But, this is not practical due to the \(\mathcal{O}(N_p N_{sn})\) complexity involved where \(N_{sn}\) is the number of nodes in the support of the kernel function. The time complexity further hides the fact that identifying the support nodes in a unstructured grid is a non-trivial task in itself. In literature, there have been a variety of ad hoc methods that have been used to project particle centered quantities to the Eulerian mesh. But these are lacking in either accuracy, consistency or are grid-dependent.
These properties become important when dealing with particle sizes about equal to or greater than the local mesh size and can severely compromise the accuracy of the simulation.
In order to sidestep the complexity issues, various alternative methods of transferring information between the particle and fluid phases have been proposed in literature.
There are different considerations for the Eulerian-to-Lagrangian and Lagrangian-to-Eulerian transfers. There are no conservation and boundedness issues when interpolating from the Eulerian mesh to the particles. It is a straight-forward interpolation and one can devise efficient schemes to do this. The only consideration is the accuracy of the interpolation.

\subsection{Particle-to-fluid transfer}
The Lagrangian-to-Eulerian transfer needs more care. For example, a particle-in-cell type approach would assign the whole volume of the particle to the Eulerian cell it is contained in. If we project using a mesh-size dependent kernel, we will end up with a mesh dependent result that can give rise to spurious and unbounded values.
The volume fraction field in particular needs to lie in \([0,1]\) to be physically meaningful. The random close packing limit for mono-disperse spheres is  $\sim0.64$. So, in reality in a non-jammed particle-laden flow the particle volume fraction will be in \([0,0.64]\).

In literature, there have been a variety of methods that have been used to project particle centered quantities to the Eulerian mesh. But these are lacking in either accuracy, consistency or are grid-dependent. For instance, \citet{casagrandeHybridFEMDEMApproach2017} use a particle-in-cell type approach where the volume occupied by a particle is assigned to the element containing the center. There are no conservation issues and  this works fine if the particles are much smaller than the mesh size. If not, there is a possibility of unphysical values appearing in the projected quantities. In \cite{elgeitaniHighOrderCFDDEMDevelopment2023b} a similar idea is used to get element-constant values that are then projected to nodal values while artificially restricting the nodal values to be within physical limits. Obtaining a volume fraction field from a particle configuration was exclusively considered in \cite{elgeitaniQuadratureCenteredAveragingScheme2023a}, where different methods are reviewed.

\begin{figure}
    \centering
    \includegraphics[width=0.5\textwidth]{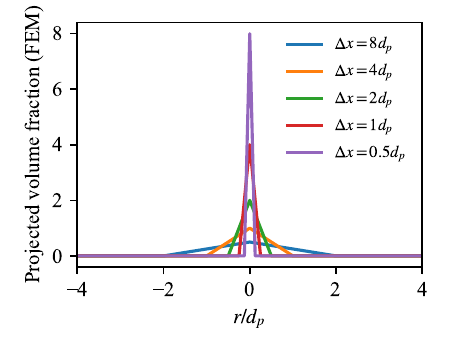}
    \caption{Lagrangian to Eulerian projection of the volume fraction for one particle centered at the origin in a 1D mesh using the FEM interpolating polynomials. As the mesh is refined, the projected value diverges\label{fig:1sdiv}}
\end{figure}

\citet{pusztayConservativeProjectionFinite2022b} use the finite element basis functions to perform the Lagrangian to Eulerian projection and directly obtain nodal values on the mesh. The process can be formulated as enforcing a weak equivalence between the particle volumes as a collection of $\delta$-functions at the particle centers and the Eulerian mesh representation of the particle volume fraction field. This is conservative, but it has same boundedness and convergence issues as the PIC-type method. In addition, depending on the specifics of the quadrature rules used, this method can generate oscillations and potentially negative values locally.
The divergent result of this method is demonstrated in \cref{fig:1sdiv}. 
It shows the projected volume fraction field on a 1D finite element mesh due to one particle at the origin under mesh refinement. As the mesh is refined, the projected value diverges.

\subsection{Two stage projection}
Here we outline a novel two-stage method that does not suffer from the issues described before and is efficient and scalable. It is a generalization of the two stage method described in \citet{Capecelatro2013} to unstructured, anisotropic and inhomogeneous meshes. The first step is along the lines of \cite{pusztayConservativeProjectionFinite2022b}, with one modification to ensure that we preserve the original sign of the quantity being projected. Let the particle distribution be denoted by 
\[\psi = \sum\limits_{i}^{N_p} V_i \delta(\bm{x}-\bm{x}_i)\]
where the $V_i$ are the particle volumes and $\bm{x}_i$ are the particle centers. The finite element representation of the volume fraction field cane be written as 
\[\phi_p = \sum\limits_{i}^{N_n} \varphi_iN_i(\bm{x})\]
where $N_i$ are the finite element basis functions and $\varphi_i$ are the nodal values. The first step is to enforce the equivalence between the particle distribution and the finite element representation of the volume fraction field in a weak sense. This is done by enforcing the following
\begin{equation}
    \int_{\Omega} \phi_p N_i(\bm{x})~\mathrm{d}\Omega = \int_{\Omega} \psi N_i(\bm{x})\mathrm{d}\Omega 
\end{equation}
This can be simplified and written as a linear system of equations of the form
\begin{equation}\label{eq:linsys_1stage}
    [M]~\{\varphi\} = [V]~\{w\}
\end{equation}
where $[M]$ is the mass matrix, $\{\varphi\}$ is the vector of nodal values and $\{w\}$ is the vector of particle volumes. $[V]$ is a sparse matrix whose rows correspond to the particle centers and columns correspond to the mesh nodes.
The mass matrix in \cref{eq:linsys_1stage} should be lumped for a couple of different reasons. First, lumping avoids oscillatory and negative values in the projected field. It also diagonalizes the linear system and the solution can be obtained by simple division at each node. This result is still mesh dependent and cannot in general be used if the particle sizes are comparable or larger than the mesh size.

% minipage for two side by side figures
\begin{figure}
    \centering
    \begin{minipage}[b]{0.45\textwidth}
        \includegraphics[width=\textwidth]{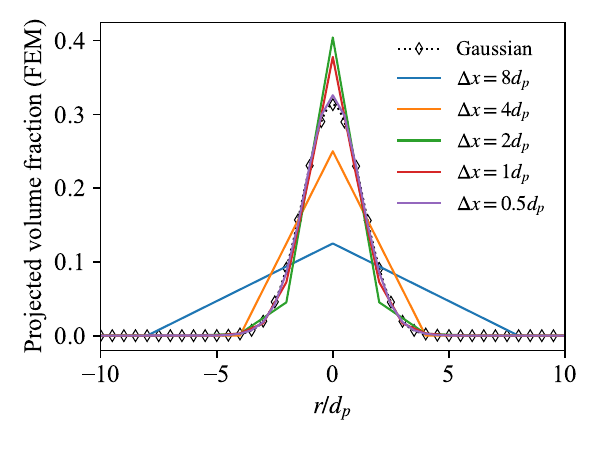}
        % \caption{Same case as before but projected using the two-stage method. As the mesh is refined, the projected value converges to a Gaussian of the prescribed width\label{fig:femconv}}
    \end{minipage}
    \hfill
    \begin{minipage}[b]{0.45\textwidth}
        \includegraphics[width=0.9\textwidth]{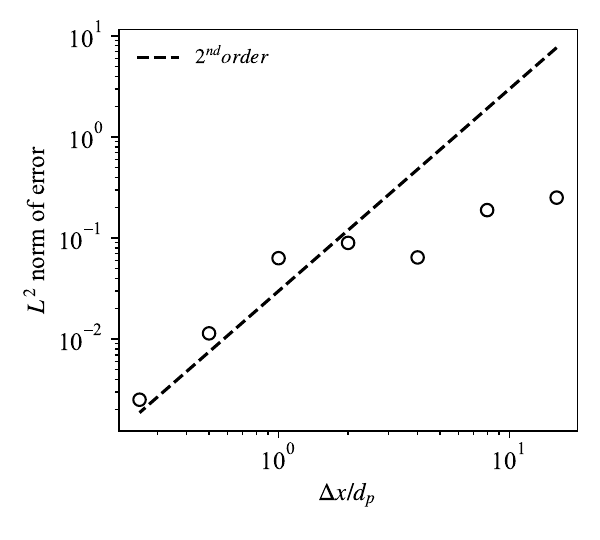}
        % \caption{Norm of error of the projected value showing second order convergence\label{fig:conv_rate}}
    \end{minipage}
    \caption{Same case as \cref{fig:1sdiv} projected using the two-stage method. Under mesh refinement, the result converges to a Gaussian of the prescribed width (i.e. filter size $\delta_f$). Right shows rate of convergence\label{fig:femconv}}
\end{figure}

The second step is to smooth this projected field using a diffusion equation, using a carefully chosen diffusivity that takes into account the local mesh size to give a mesh independent smoothing effect. The diffusion equation is discretized using the same finite elements and advanced using pseudo-time-stepping on the same domain as the fluid.
This can be written as 
\begin{equation}\label{eq:l2estage2}
    \pdv{\phi_p}{\tau} = \nabla\cdot\left(\mathscr{D}\nabla\phi_p\right)
\end{equation}
with a diffusivity defined as 
\begin{equation}
    \mathscr{D} = \max{\left(\frac{\delta_f^2-\Delta x^2}{16\ln 2},0\right)}
\end{equation}
where $\delta_f$ is the filter size/smoothing-length-scale set to be $4 d_p$ and $\Delta x$ is a nominal local mesh size that is calculated as the average of the mesh edge lengths for each node.
Here, $\tau$ is the pseudo time variable and \cref{eq:l2estage2} is advanced from $\tau=0$ to $\tau=1$. This corresponds to a smoothing length scale of $\sqrt{\mathscr{D}\tau}$ which corresponds to a Gaussian with a full width at half maximum of $\delta_f$ i.e. it is equivalent to analytically transforming a Dirac delta function to a Gaussian of width $\delta_f$ while conserving the integrated value. While the method is demonstrated here on a 1D case, the extension to higher dimensions is quite straight forward and we have verified that the 3D version retains the desired  properties.

\section{Example cases}
The method developed here is applied to a few different particle-laden flow cases. This is preliminary work and the focus is on demonstrating the capabilities of the implementation. The results presented here are presently only qualitative. In addition to a simple pipe geometry, we demonstrate the capabilities of the method by simulating particles flowing through a simplified bifurcation geometry that one might find in the arterial vasculature.

\subsection{Pressure driven flow in a pipe}
\begin{figure}
    \centering
    \includegraphics[width=0.9\textwidth]{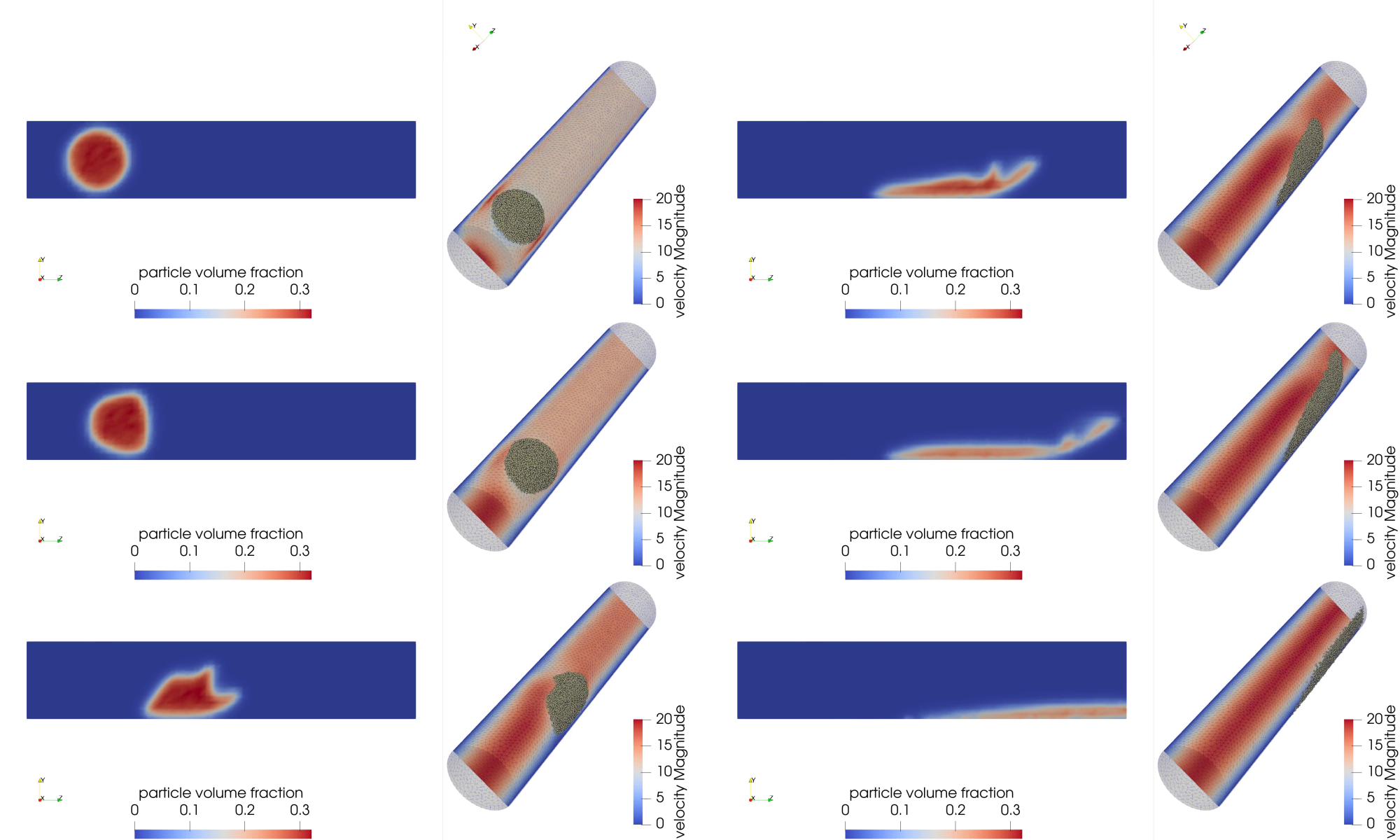}
    \caption{Fully coupled simulation of particles in a pressure driven pipe flow. Results at different instances are shown. In each frame, the figure on the left shows contour plot of the volume fraction of the particles on a vertical slice through the center of the cylinder. The figure on the right shows a contour plot of the magnitude of the velocity field along with the particles themselves. }
    \label{fig:pipe_flow}
\end{figure}

The first case considered is pressure driven flow in a cylindrical tube. The geometry is shown in figure \ref{fig:pipe_flow}. The Reynolds number based on the pipe diameter is 100. There are 10,000 particles are initialized in as a spherical bolus near the inlet boundary. The domain is discretized using 24,000 elements. The inlet velocity is specified as parabolic flow profile. The outlet is a zero pressure boundary condition and the walls are no-slip boundaries. The particle are denser than the carrier fluid, $\rho_p = 5\rho_f$ and gravity is included. This case is solved in parallel with 4 MPI processes.

In figure~\ref{fig:pipe_flow} we see that the particles are transported down the cylinder by the fluid. Due to the presence of gravity particles also settle along the bottom. The fully-coupled nature of the simulation is evident from the velocity contours. Initially, the particles are at rest and block the fluid. As a result, he fluid has to accelerate around the particles. This delays the development of the parabolic profile downstream of the particles. The particles slowly accelerate due to the fluid and gradually pick up speed and are carried out of the cylinder through the outlet.

\subsection{Flow in bifurcation geometry}
\subsubsection*{Flow in a simple bifurcation}
\begin{figure}
    \centering
    \includegraphics[width=0.9\textwidth]{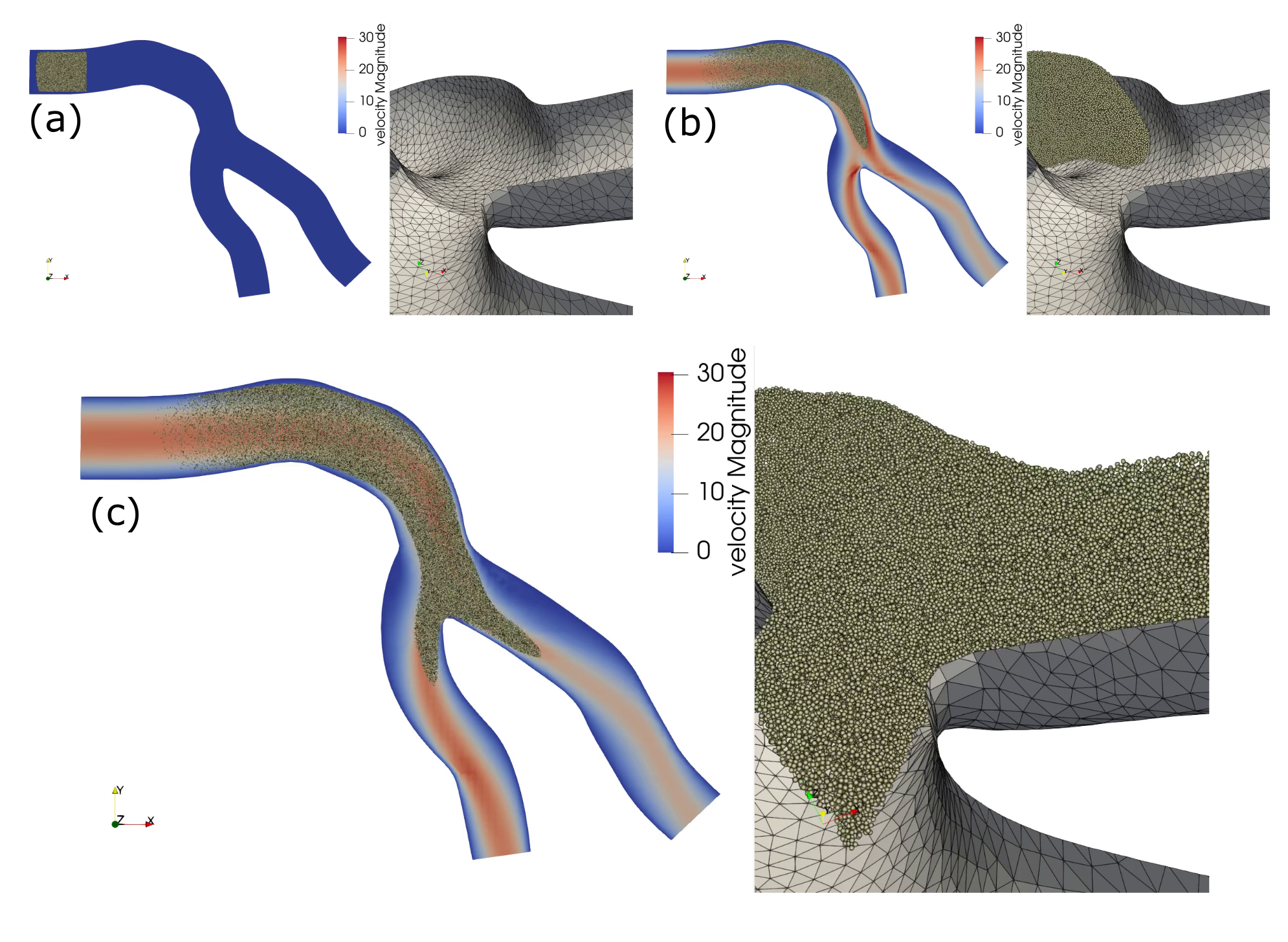}
    \caption{Fully coupled simulation of particles flowing through a bifurcation. The initial condition is shown in (a). Snapshots of the velocity contour and particle locations at subsequent times are are shown in (b) and (c)\label{fig:bifurcation}}
\end{figure}

Another example demonstrated here is particle-laden flow through a bifurcation geometry shown in \cref{fig:bifurcation}. The fluid domain is vessel that splits into two smaller vessels, that resembles a typical feature of the arterial vasculature. The vessel geometry is discretized using 220K tetrahedral elements. We initialize 100,000 particles in the inlet in a cylindrical bolus with a mean particle volume fraction in the bolus of $\sim 0.35$. The particles and fluid are initially at rest. The inlet velocity is specified with a parabolic profile corresponding to a Reynolds number of 250 based on the inlet flow and diameter. There is no gravity and the particles density is twice that of the fluid. The outlet vessels are coupled to a 0D lumped parameter model that models the downstream vasculature. The lumped parameter model is a 3-element Windkessel model. Since this is a steady flow simulation, after the lumped parameter capacitor is charged, the outflow boundary conditions reduce to outlets with a resistance.

\Cref{fig:bifurcation} shows the initial condition in the top left panel and snapshots at two subsequent instances as the simulation progresses. Within each subfigure, the left panel shows a contour plot of the velocity field magnitude along with the particle positions. The right panel is a close-up view of the bifurcation region looking up stream into the inlet vessel. 
As the particles come to the bifurcation, they collide with the boundary walls and get split into both the downstream vessels following complex inertial trajectories.

\subsubsection*{Flow in a bifurcation with a stenosis}
\begin{figure}
    \centering
    \includegraphics[width=0.95\textwidth]{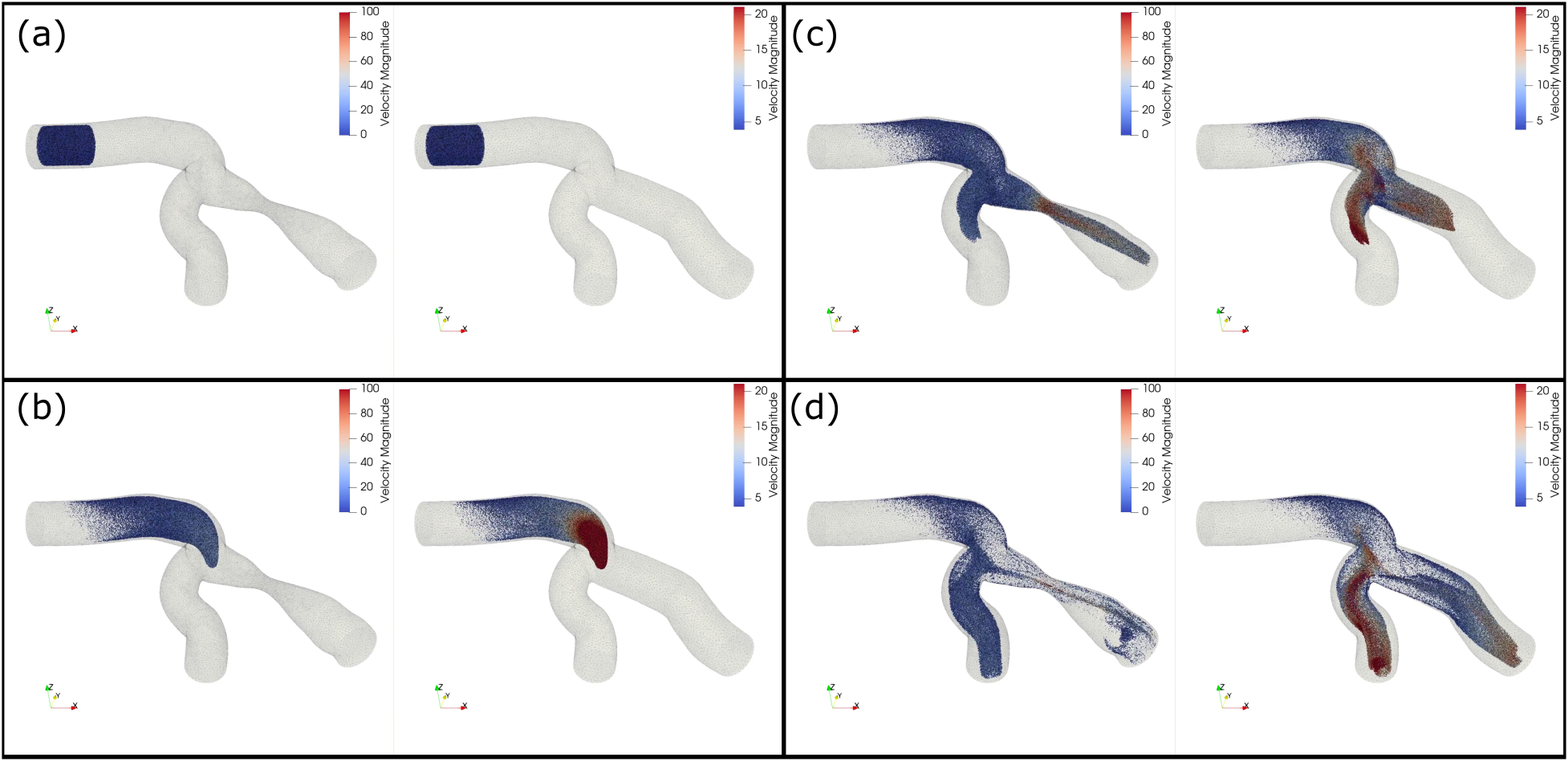}
    \caption{ Comparison of particles flowing through a bifurcation with and without a stenosis. In (d) the particles can be seen to be flowing back away from the outlet in the stenosed branch. This is due to the formation of a circulating flow topology caused by the stenosis.\label{fig:bifur_stenosed_comp}}
\end{figure}

The final example shown is a modification of the bifurcation geometry from the previous example. Here, we introduce a stenosis in one of the daughter vessels, to simulate a narrowing of the vessel due to atherosclerosis. The rest of the conditions are unchanged. The results for this case are shown in \cref{fig:bifur_stenosed_comp}.

Within each subfigure in \cref{fig:bifur_stenosed_comp}, the previous healthy geometry is on the right and the stenosed geometry is on the left. We compare these two simulations at same instances in time. Till subfigure (b) the the particle configurations do not show any significant difference between the two cases. In \cref{fig:bifur_stenosed_comp}c we see that the particles are accelerated through the narrowing and reach the outlet much quicker than before. In the final subfigure, in the stenosed geometry we see a particles form a spur-like shape and flowing backward. This is due to a recirculation zone that forms downstream of the stenosis, a well known feature of flow through stenosed vessels. 

These examples are meant to be a qualitative demonstration of the capabilities of the method. In the future, this work will be followed up with validation and performance evaluation and scaling studies. We will also apply this method to more larger and complex geometries derived from patient-specific medical images.

\section{Conclusions}
To summarize, we have presented a computational framework for modeling large scale particle-laden flows in complex geometries that enable subject-specific analysis. The framework is based on a volume-filtered Euler-Lagrange approach that uses a finite element method for the fluid phase and a discrete element method for the particle phase. The fluid phase is solved on a three-dimensional unstructured grid using a stabilized finite element method. The particle phase is modeled as rigid spheres and their motion is calculated according to Newton's second law for translation and rotation. The hydrodynamic force on the particles is calculated using a recently developed correlation for freely evolving suspensions of particles. The method is applied to a few different particle-laden flow cases. The results are of a qualitative nature and are not meant to be quantitative. The results demonstrate the capabilities of the implementation and the potential of the method for simulating large-scale particle-laden flows in complex geometries.
We intend to follow up this work with validation and performance evaluation and scaling studies. We will also apply this method to more larger and complex geometries derived from patient-specific medical images and other biofluids applications.
\bibliography{vfel_fem_dem}
\end{document}